\documentclass[final,5p,times,twocolumn]{elsarticle}

\usepackage{bm}
\usepackage{bbm}
\usepackage{bbold}
\usepackage{slashed}
\usepackage{verbatim}
\usepackage{cancel}
\usepackage{graphicx}
\usepackage{combelow} 
\usepackage{mathtools}
\usepackage[usenames,dvipsnames,svgnames,table]{xcolor}
\usepackage[colorlinks=True,linkcolor=blue,citecolor=blue,urlcolor=blue]{hyperref}
\usepackage{soul}
\usepackage{ulem}

\let\oldciteauthor=\citeauthor
\def\citeauthor#1{\hypersetup{citecolor=black}\oldciteauthor{#1}}

\let\oldciten=\onlinecite
\def\onlinecite#1{\hypersetup{citecolor=blue}\oldciten{#1}}

\let\oldcite=\cite
\def\cite#1{\hypersetup{citecolor=blue}\oldcite{#1}}

\makeatletter
\newsavebox{\@brx}
\newcommand{\llangle}[1][]{\savebox{\@brx}{\(\m@th{#1\langle}\)}%
  \mathopen{\copy\@brx\kern-0.5\wd\@brx\usebox{\@brx}}}
\newcommand{\rrangle}[1][]{\savebox{\@brx}{\(\m@th{#1\rangle}\)}%
  \mathclose{\copy\@brx\kern-0.5\wd\@brx\usebox{\@brx}}}
\makeatother

\newcommand{\beqn}{\begin{eqnarray}}
\newcommand{\eeqn}{\end{eqnarray}}
\newcommand{\beqs}{\begin{subequations}}
\newcommand{\eeqs}{\end{subequations}\\[-2mm]\noindent}
\newcommand{\eq}[1]{(\ref{#1})}

\newcommand{\Z}{{\mathbb Z}}

\newcommand{\bs}{\boldsymbol}

\definecolor{purple}{rgb}{0.8,0,0.6}
\definecolor{PURPLE}{rgb}{0.8,0,0.6}
\definecolor{orange}{rgb}{1,0.55,0}
\definecolor{limegreen}{rgb}{0.2,0.8,0.2}

\allowdisplaybreaks

\journal{Physics Letters B}

\begin{document}

\begin{frontmatter}

\title{Acceleration as a circular motion along an imaginary circle: Kubo-Martin-Schwinger condition for accelerating field theories in imaginary-time formalism}

\author[uvt]{Victor E. Ambru\cb{s}}
\ead{victor.ambrus@e-uvt.ro}
\author[idp,uvt]{Maxim N. Chernodub}
\ead{maxim.chernodub@idpoisson.fr}
\address[uvt]{Department of Physics, West University of Timi\cb{s}oara,\\
Bd.~Vasile P\^arvan 4, Timi\cb{s}oara 300223, Romania}
\address[idp]{Institut Denis Poisson, Universit\'e de Tours, Tours 37200, France}

\begin{abstract}
We discuss the imaginary-time formalism for field theories in thermal equilibrium in uniformly accelerating frames. We show that under a Wick rotation of Minkowski spacetime, the Rindler event horizon shrinks to a point in a two-dimensional subspace tangential to the acceleration direction and the imaginary time. We demonstrate that the accelerated version of the Kubo-Martin-Schwinger (KMS) condition implies an identification of all spacetime points related by integer-multiple rotations in the tangential subspace about this Euclidean Rindler event-horizon point, with the rotational quanta defined by the thermal acceleration, $\alpha = a/T$. In the Wick-rotated Rindler hyperbolic coordinates, the KMS relations reduce to standard (anti-)periodic boundary conditions in terms of the imaginary proper time (rapidity) coordinate. Our findings pave the way to study, using first-principle lattice simulations, the Hawking-Unruh radiation in geometries with event horizons, phase transitions in accelerating Early Universe and early stages of quark-gluon plasma created in relativistic heavy-ion collisions.
\end{abstract}

\begin{keyword}
Acceleration \sep 
Unruh effect \sep
KMS relation \sep 
Finite temperature field theory
\end{keyword}
\end{frontmatter}

\date{\today}

\section{Introduction}\label{sec:intro}

In the past decades, there has been a renewed interest in studying systems with acceleration as toy models for understanding the dynamics of the quark-gluon plasma fireball created in ultrarelativistic (non-central) heavy-ion collisions~\cite{Castorina:2007eb}. Such systems exhibit large acceleration immediately after the collision \cite{Kharzeev:2005iz} until the central rapidity plateau develops as in the Bj\"orken boost-invariant flow model~\cite{Bjorken:1982qr}, where the acceleration vanishes. A natural question that arises for such a system is to what extent these extreme kinematic regimes affect the thermodynamics of the plasma fireball, which sets the stage for further evolution of the quark-gluon plasma. The environment of the ``Little Bangs'' of high-energy heavy-ion collisions~\cite{Gelis2016} sheds insights on the properties of a primordial quark-gluon matter that once emerged at the time of the Big Bang in the Early Universe~\cite{yagi2008quark}.

Our knowledge of the non-perturbative properties of the quark-gluon plasma originates from first-principle numerical simulations of lattice QCD, which is formulated in Euclidean spacetime, by means of the imaginary-time formalism~\cite{Kapusta:2006pm}. Acceleration is closely related to rotation due to the resemblance of the corresponding generators of Lorentz transformations of Minkowski spacetime. In the case of non-central collisions, the angular velocity of the quark-gluon fluid can reach values of the order of $\Omega \sim 10^{22}\,{\rm Hz}$~\cite{STAR:2017ckg} which translates to $\hbar \Omega \simeq 6\ {\rm MeV} \ll T_c$, where $T_c$ is the transition temperature to the quark-gluon plasma phase. The lattice studies have so far been limited to the case of uniformly rotating systems in Euclidean space-time, where the rotation parameter has to be analytically continued to imaginary values~\cite{Yamamoto:2013zwa} in order to avoid the sign problem that also plagues lattice calculations at finite chemical potential~\cite{deForcrand:2002hgr}. Analytical analyses of the effects of rotation on the phase diagram, performed in various effective infrared models of QCD~\cite{Chen:2015hfc,Jiang:2016wvv,Chernodub:2016kxh,Wang:2018sur,Sadooghi:2021upd,Chen:2020ath,Fujimoto:2021xix,Chernodub:2020qah}, stay in persistent contradiction with the first-principle numerical results~\cite{Braguta:2020biu,Braguta:2021jgn,Braguta:2022str,Braguta:2023yjn,Braguta:2023kwl,Yang:2023vsw}, presumably due to numerically-observed rotational instability of quark-gluon plasma~\cite{Braguta:2023yjn,Braguta:2023kwl,Yang:2023vsw} (related to the thermal melting of the non-perturbative gluon condensate~\cite{Braguta:2023yjn}), splitting of chiral and deconfining transitions~\cite{Yang:2023vsw,Sun:2023dwh}, or formation of a strongly inhomogeneous mixed hadronic--quark-gluon-plasma phase induced by rotation~\cite{Chernodub:2020qah,Chernodub:2022veq}.

An earlier study of a Euclidean quantum field theory in an accelerating spacetime with the Friedmann-Lema\^itre-Robertson-Walker metric has also encountered the sign problem, which was avoided by considering a purely imaginary Hubble constant~\cite{Yamamoto:2014vda}. On the contrary, our formulation of acceleration in the imaginary-time formalism is free from the sign problem, and thus, it can be formulated for physical, real-valued acceleration. Throughout the paper, we use $\hbar = c = k_B = 1$ units.

\section{Global equilibrium in uniform acceleration}
From a classical point of view, global equilibrium states in generic particle systems are characterized by the inverse temperature four-vector $\beta^\mu \equiv u^\mu(x) / T(x)$, associated with the local fluid velocity $u^\mu$, with $\beta^\mu$ satisfying the Killing equation, $\partial_\mu \beta_\nu + \partial_\nu \beta_\mu = 0$ \cite{Cercignani:2002,Becattini:2012tc}. For an accelerated system at equilibrium, one gets $\beta^\mu \partial_\mu = \beta_T [(1 + a z) \partial_t + a t \partial_z]$, with $\beta_T = 1/T$ where\footnote{Throughout our article, $T(x)$ denotes the local temperature~\eq{eq_T_z}, while $T$ stands for the value of $T(x)$ at the origin $t = z = 0$. Also, for reasons that will become clear shortly later, we use the notation $\beta_T$ instead of the conventional $\beta$ for the inverse temperature at the coordinate origin}. $T \equiv T({\bs 0})$ represents the temperature at the coordinate origin ${\bs x}_{\|} \equiv (t,z) = {\bs 0}$ in the longitudinal plane spanned by the time coordinate $t$ and the acceleration direction $z$. The local temperature $T(x)$, the local fluid velocity $u^\mu(x)$ and the local proper acceleration $a^\mu(x) \equiv u^\nu \partial_\nu u^\mu$, respectively,
\begin{align}
 T(x) &\equiv (u_\mu \beta^\mu)^{-1} = \frac{1}{\beta_T \sqrt{(1 + a z)^2 - (a t)^2}},
\label{eq_T_z} \\ 
u^\mu(x) \partial_\mu & = T(x) \beta_T 
\bigl[(1 + az) \partial_t + a t \partial_z\bigr]\,,
	\label{eq_u}\\
    a^\mu(x) \partial_\mu &= a T^2(x) \beta_T^2 [at \partial_t + (1 + az) \partial_z]\,,
\label{eq_a}
\end{align}
diverge at the Rindler horizon:
\begin{align}
    (1 + az)^2 - (at)^2 = 0, \qquad\ z \geqslant - \frac{1}{a}\,.
\label{eq_Rindler_horizon_M}
\end{align}

It is convenient to define the dimensionless quantity called the proper thermal acceleration $\alpha = \sqrt{-\alpha^\mu \alpha_\mu}$ and the corresponding four-vector $\alpha^\mu = u^\nu \partial_\nu \beta^\mu = a^\mu / T(x)$, respectively:
\begin{align}
    \alpha &= a \beta_T\,, & 
    \alpha^\mu(x) \partial_\mu &= a \beta_T^2 T(x) [a t \partial_t + (1 + az) \partial_z]\,.
    \label{eq_alpha}
\end{align}
Note that, while the magnitude $\alpha$ of the thermal acceleration is a space-time constant, the local acceleration $a(x) = \sqrt{-a_\mu a^\mu} = \alpha T(x)$ depends on space and time coordinates.

In classical theory, the energy-momentum tensor for an accelerating fluid in thermal equilibrium reads
\begin{equation}
 T^{\mu\nu} = \mathcal{E} u^\mu u^\nu - \mathcal{P} \Delta^{\mu\nu},
 \label{eq_Tmunu}
\end{equation}
where $\Delta^{\mu\nu} = g^{\mu\nu} - u^\mu u^\nu$.
The local energy density $\mathcal{E}$ and pressure $\mathcal{P}$ are characterized by the local temperature~\eq{eq_T_z}. For a conformal system, 
\begin{equation}
 \mathcal{E} = 3\mathcal{P} = \frac{\nu_{\rm eff} \pi^2}{30} T^4(x),
 \label{eq_eps}
\end{equation}
where $\nu_{\rm eff}$ is the effective bosonic degrees of freedom. In the case of a massless, neutral scalar field, $\nu_{\rm eff} = 1$, while for Dirac fermions, $\nu_{\rm eff} = \frac{7}{8} \times 2 \times 2 = 7/2$, taking into account the difference between Bose-Einstein and Fermi-Dirac statistics ($7/8$), spin degeneracy, as well as particle and anti-particle contributions.

\section{Unruh and Hawking effects}
Unruh has found that in a frame subjected to a uniform acceleration $a$, an observer detects a thermal radiation with the temperature~\cite{Unruh:1976db}:
\begin{align}
 T_U \equiv \frac{1}{\beta_U} = \frac{a}{2\pi}\,,
\label{eq_T_U}
\end{align}
where we also defined the Unruh length $\beta_U$, which will be useful in our discussions below.

The Unruh effect is closely related to the Hawking evaporation of black holes~\cite{Hawking1974,Hawking1975}, which proceeds via the quantum production of particle pairs near the event horizon of the black hole. The Hawking radiation has a thermal spectrum with an effective temperature
\begin{align}
	T_H = \frac{\kappa}{2\pi}\,,
\label{eq_T_H}
\end{align}
where $\kappa=1/(4M)$ is the acceleration due to gravity at the horizon of a black hole of mass $M$. The similarity of both effects, suggested by the equivalence of formulas for the Unruh temperature~\eq{eq_T_H} and the  Hawking temperature~\eq{eq_T_U}, goes deeper as the thermal character of both phenomena apparently originates from the presence of appropriate event horizons~\cite{Gibbons:1976pt,Gibbons:1976es}. In an accelerating frame, the event horizon separates causally disconnected regions of spacetime, evident in the Rindler coordinates in which the metric of the accelerating frame is conformally flat~\cite{Birrell:1982ix}.

Quantum effects lead to acceleration-dependent corrections to Eq.~\eqref{eq_eps} and may also produce extra (anisotropic) contributions to the energy-momentum tensor $T^{\mu\nu}$ of the system. Such corrections were already established using the Zubarev approach \cite{Prokhorov:2019cik,Prokhorov:2019sss} or Wigner function formalism \cite{Becattini:2020qol,Palermo:2021hlf}, and one remarkable conclusion is that the energy-momentum tensor $\Theta^{\mu\nu}$ in an accelerating system exactly vanishes at the Unruh temperature~\eq{eq_T_U}, or, equivalently, when the thermal acceleration~\eq{eq_a} reaches the critical value $\alpha = \alpha_c = 2\pi$: $\Theta^{\mu\nu}(T = T_U) = 0$. A somewhat related property is satisfied by thermal correlation functions in the background of a Schwarzschild black hole, establishing the equivalence between Feynman and thermal Green's functions, with the latter one taken at the Hawking temperature~\eq{eq_T_H}, cf.~Ref.~\cite{Gibbons:1976es,Gibbons:1976pt}. 

As noted earlier, the energy density receives quantum corrections. For the conformally-coupled massless real-valued Klein-Gordon scalar field and the Dirac field, we have, respectively~\cite{Prokhorov:2019sss,Becattini:2020qol,Palermo:2021hlf,Ambrus:2014itg,Ambrus:2021eod}:
\begin{subequations}\label{eq_E}
\begin{align}
\mathcal{E}_{\rm scalar} &= \frac{\pi^2 T^4(x)}{30} \biggl[1 - \Bigr(\frac{\alpha}{2\pi}\Bigl)^4\biggr]\,,\\
\mathcal{E}_{\rm Dirac} & = \frac{7\pi^2 T^4(x)}{60} \biggl[1 - \Bigr(\frac{\alpha}{2\pi}\Bigl)^2\biggr] \biggl[1 + \frac{17}{7} \Bigr(\frac{\alpha}{2\pi}\Bigl)^2\biggr]\,,
\end{align}
\end{subequations}
where we specially rearranged terms to make it evident that at the Unruh temperature $T = T_U$ (or, equivalently, at $\alpha = 2\pi$), the energy density vanishes.

The above discussion focused on the free-field theory. In the interacting case, a legitimate question is to what extent do the local kinematics influence the phase structure of phenomenologically relevant field theories, for example, to deconfinement and chiral thermal transitions of QCD. Central to lattice finite-temperature studies is how to set the Euclidean-space boundary conditions in the imaginary-time formalism. A static bosonic (fermionic) system at finite temperature can be implemented by imposing (anti-)periodicity 
in imaginary time $\tau = i t$ with period given by the inverse temperature, $\tau \rightarrow \tau + \beta_T$. These boundary conditions are closely related to, and in fact, derived from the usual Kubo-Martin-Schwinger (KMS) relation formulated for a finite-temperature state (at vanishing acceleration), which translates into a condition written for the scalar and fermionic thermal two-point functions \cite{Kapusta:2006pm,Mallik:2016anp}:
\begin{equation}
 G_F(t) = G_F(t + i\beta_T), \quad 
 S_F(t) = -S_F(t + i\beta_T),
 \label{eq_standard_KMS}
\end{equation}
where we suppressed the dependence on the spatial coordinate $\bs x$ and the second four-point $x'$.
In the case of rotating states, the KMS relation~\eq{eq_standard_KMS} gets modified to~\cite{Chernodub:2020qah,Ambrus:2021eod,Ambrus:2019gkt}
\begin{align}
 G_F(t, \varphi) &= G_F(t + i\beta_T, \varphi + i \beta_T \Omega),\nonumber\\
 S_F(t,\varphi) &= - e^{-\beta_T \Omega S^z} S_F(t + i \beta_T, \varphi + i \beta_T \Omega),
\end{align}
where $e^{-\beta_T \Omega S^z}$ is the spin part of the rotation with imaginary angle $i \beta_T \Omega$ along the rotation ($z$) axis and $S^z = \frac{i}{2} \gamma^x \gamma^y$ is the spin matrix. The purpose of the present paper is to uncover the KMS relation and subsequent conditions for fields and, consequently, for correlation functions in a uniformly accelerated state. 

\section{Quantum field theory at constant acceleration} 
In Minkowski space, the most general solution of the Killing equation reads
\begin{equation}
 \beta^\mu = b^\mu + \varpi^\mu{}_\nu x^\nu,
\end{equation} 
where $b^\mu$ is a constant four-vector and $\varpi^{\mu\nu}$ is a constant, anti-symmetric tensor. A quantum system in thermal equilibrium is characterized by the density operator
\begin{equation}
 \hat{\rho} = e^{-b \cdot \hat{P} + \varpi : \hat{J} /2},
 \label{eq:rho_GTE}
\end{equation}
where $\hat{P}^\mu$ and $\hat{J}^{\mu\nu}$ are the conserved four-momentum and total angular momentum operator,
representing the generators of translations and of Lorentz transformations.
In order to derive the KMS relation, it is convenient to factorize $\hat{\rho}$ into a translation part and a Lorentz transformation part, as pointed out in Ref.~\cite{Becattini:2020qol}:
\begin{equation}
 e^{-b \cdot \hat{P} + \varpi : \hat{J} / 2} = e^{-\tilde{b}(\varpi) \cdot \hat{P}} e^{\varpi : \hat{J} / 2},
\end{equation}
where $\tilde{b}$ is given by
\begin{equation}
 \tilde{b}(\varpi)^\mu = \sum_{k= 0}^\infty \frac{i^k}{(k+1)!} (\varpi^\mu{}_{\nu_1} \varpi^{\nu_1}{}_{\nu_2}  \cdots \varpi^{\nu_{k-1}}{}_{\nu_k}) b^{\nu_k}.
\end{equation}

Focusing now on the accelerated system with reference inverse temperature $\beta_T = 1/T$, we have $b^\mu = \beta_T \delta^\mu_0$ and $\varpi^\mu{}_\nu = \alpha (\delta^\mu_3 g_{0\nu} - \delta^\mu_0 g_{3\nu})$, such that $\tilde{b}$ becomes
\begin{equation}
 \tilde{b}^\mu = B \delta^\mu_0 + A \delta^\mu_3, \quad 
 B = \frac{\sin \alpha}{a}, \quad 
 A = \frac{i}{a}(1 - \cos \alpha),
\end{equation}
where $\alpha = a/T$ is the thermal acceleration~\eq{eq_alpha}.
This observation allows $\hat{\rho} = e^{-\beta_T \hat{H} + \alpha \hat{K}^z}$ to be factorized as
\begin{equation}
 \hat{\rho} = e^{-B \hat{H} + A \hat{P}^z} e^{\alpha \hat{K}^z}.\label{eq_rho_fact}
\end{equation}

A relativistic quantum field described by the field operator $\hat{\Phi}$ transforms under Poincar\'e transformations as
\begin{align}
 e^{i \tilde{b} \cdot \hat{P}} \hat{\Phi}(x) e^{-i \tilde{b} \cdot \hat{P}} &= \hat{\Phi}(x + \tilde{b}), \nonumber\\
 \hat{\Lambda} \hat{\Phi}(x) \hat{\Lambda}^{-1} &= D[\Lambda^{-1}] \hat{\Phi}(\Lambda x),\label{eq_Poincare}
\end{align}
where $\Lambda = e^{-\frac{i}{2} \varpi : \mathcal{J}}$ is written in terms of the matrix generators $(\mathcal{J}^{\mu\nu})_{\alpha \beta} = i(\delta^\mu_\alpha \delta^\nu_\beta - \delta^\mu_\beta \delta^\nu_\alpha)$, while $D[\Lambda]^{-1} = e^{\frac{i}{2} \varpi:S}$ is the spin part of the inverse Lorentz transformation. 
Comparing Eq.~\eqref{eq_Poincare} and \eqref{eq:rho_GTE}, it can be seen that the density operator $\hat{\rho}$ acts like a Poincar\'e transformation with imaginary parameters \cite{Becattini:2020qol}.
Using now the factorization \eqref{eq_rho_fact}, 
it can be seen that $\hat{\rho}$ acts on the field operator $\hat{\Phi}$ as follows:
\begin{equation}
 \hat{\rho} \hat{\Phi}(t, z) \hat{\rho}^{-1}
 = e^{-\alpha S^{0z}} \hat{\Phi}({\tilde  t}, {\tilde  z}),
\end{equation}
where
\begin{align}
 {\tilde  t} &= \cos(\alpha) t + i \sin( \alpha) z + \frac{i}{a} \sin(\alpha), \nonumber\\
 {\tilde  z} &= i \sin( \alpha) t + \cos( \alpha) z - \frac{1}{a} [1 - \cos (\alpha)].
 \label{eq_Minkowski_tz}
\end{align}
The spin term evaluates to $e^{- \alpha S^{0z}} = 1$ in the scalar case (since $S^{0z} = 0$), while for the Dirac field, $S^{0z} = \frac{i}{2} \gamma^0 \gamma^3$ and
\begin{equation}
 e^{- \alpha S^{0z}} = \cos \frac{\alpha}{2} - i \gamma^0 \gamma^3 \sin\frac{\alpha}{2}. \label{eq22}
\end{equation}

\section{KMS relation at constant uniform acceleration} 
Consider now the Wightman functions $G^{\pm}(x,x')$ and $S^{\pm}(x,x')$ of the Klein-Gordon and Dirac theories, defined respectively as
\begin{align}
 G^+(x,x') &= \langle \hat{\Phi}(x) \hat{\Phi}(x') \rangle, & 
 S^+(x,x') &= \langle \hat{\Psi}(x) \hat{\overline{\Psi}}(x') \rangle, \nonumber\\
 G^-(x,x') &= \langle \hat{\Phi}(x') \hat{\Phi}(x) \rangle, &
 S^-(x,x') &= -\langle \hat{\overline{\Psi}}(x') \hat{\Psi}(x) \rangle.
\end{align}
When the expectation value $\langle \cdot \rangle$ is taken at finite temperature and under acceleration, we derive the KMS relations:
\begin{align}
 G^+(x,x') & = G^-({\tilde  t}, {\tilde  z};x'), \nonumber\\
 S^+(x,x') &= -e^{-\alpha S^{0z}} S^-({\tilde  t}, {\tilde  z};x').
\end{align}

The KMS relations also imply natural boundary conditions for the thermal propagators:
\begin{align}
 G_F({\tilde  t}, {\tilde  z}; x') &= G_F(t, z; x')\,, \nonumber\\ 
 S_F({\tilde  t}, {\tilde  z}; x') &= - e^{\alpha S^{0z}} S_F(t, z; x')\,,
\label{eq_acc_bc}
\end{align}
which are solved formally by \cite{Birrell:1982ix,Ambrus:2021eod}
\begin{subequations}\label{eq_thermal_Feynman}
\begin{align}
 G_F^{(\alpha)}(t, z; x') &= \sum_{j = -\infty}^\infty G_F^{\rm vac}(t_{(j)}, z_{(j)}; x')\,, \\
 S_F^{(\alpha)}(t, z; x') &= \sum_{j = -\infty}^\infty (-1)^j e^{-j \alpha S^{0z}} S_F^{\rm vac}(t_{(j)}, z_{(j)}; x')\,,
\end{align}
\end{subequations}
where $G^{\rm vac}_F(x,x')$ and $S^{\rm vac}_F(x,x')$ are the vacuum propagators, while $t_{(j)}$ and $z_{(j)}$ are obtained by applying the transformation in Eq.~\eqref{eq_Minkowski_tz} $j \in \Z$ times:
\begin{align}
 t_{(j)} &= t \cos(j \alpha) + \frac{i}{a} (1 + az) \sin(j \alpha), \nonumber\\
 z_{(j)} &= it \sin(j \alpha) + \frac{1}{a} (1 + a z) \cos(j \alpha) - \frac{1}{a}.
 \label{eq_Minkowski_tzj}
\end{align}
In particular, $\tilde{t} = t_{(1)}$ and $\tilde{z} = z_{(1)}$. Due to the periodicity of the trigonometric functions appearing above, in the case when $\alpha / 2\pi = p/q$ is a rational number represented as an irreducible fraction, the sum over $j$ in Eqs.~\eqref{eq_thermal_Feynman} contains only $q$ terms:
\begin{subequations}\label{eq_thermal_Feynman_rational}
\begin{align}
 G_F^{(p,q)}(t, z; x') &= \sum_{j = 0}^{q-1} G_F^{\rm vac}(t_{(j)}, z_{(j)}; x'), \\
 S_F^{(p,q)}(t, z; x') &= \sum_{j = 0}^{q-1} (-1)^j e^{-j \alpha S^{0z}} S_F^{\rm vac}(t_{(j)}, z_{(j)}; x').
\end{align}
\end{subequations}
In particular, the case $\alpha = 2\pi$ corresponds to $p = q = 1$, while the thermal propagators reduce trivially to the vacuum ones: $G_F^{(1,1)} = G_F^{\rm vac}$ and $S_F^{(1,1)} = S_F^{\rm vac}$.
Since $e^{-q \alpha S^{0z}} = (-1)^p$ by virtue of Eq.~\eqref{eq22}, applying Eq.~\eqref{eq_acc_bc} $q$ times shows that $S_F^{(p,q)}(t_{(q)}, z_{(q)}; x') = (-1)^{p+q} S^{(p,q)}_F(t, z; x')$ and thus $S_F^{(p,q)}$ cancels identically when $p + q$ is an odd integer.

\section{Imaginary-time formulation for acceleration} 

\begin{figure}[!ht]
\begin{center}
\includegraphics[width=0.45\textwidth,clip=true]{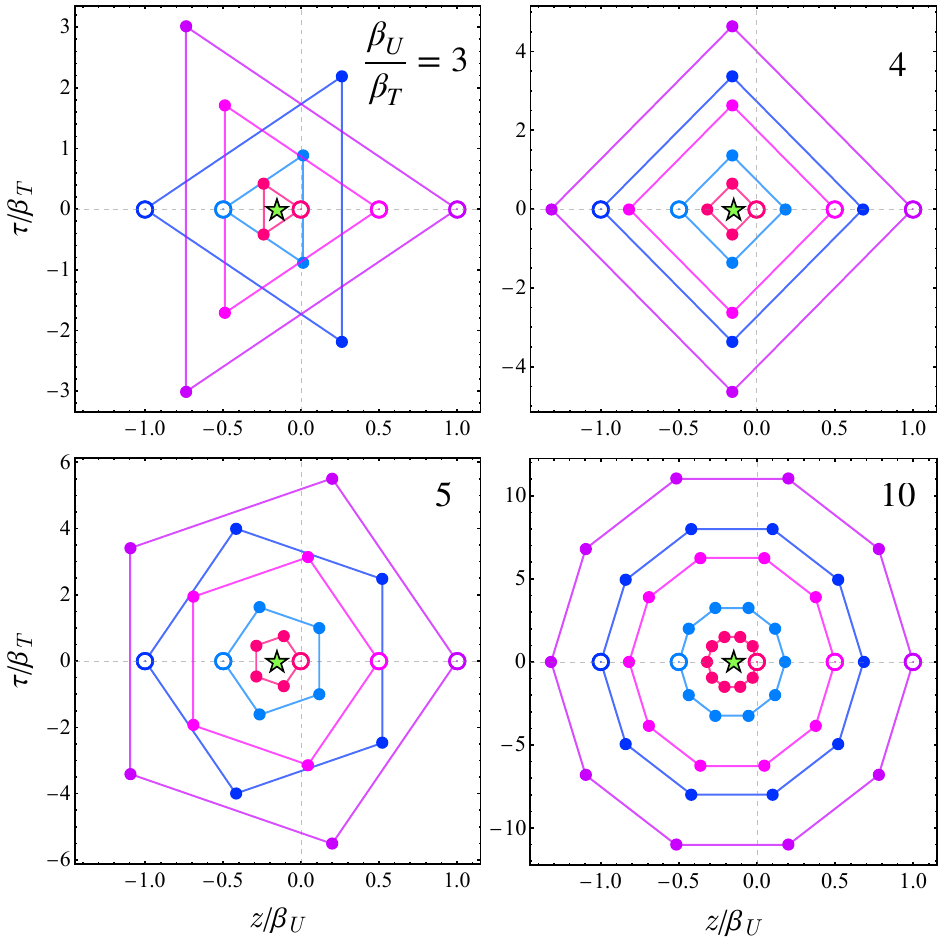}
\end{center}
\caption{
The cyclic paths determined by the accelerating KMS boundary condition~\eq{eq_Euclid_tau_zj_0} in the longitudinal plane spanned by the imaginary time $\tau$ and the acceleration direction $z$ of Wick-rotated Minkowski spacetime. Each plot illustrates different accelerations $a$ encoded in the ratio $\beta_U/\beta_T \equiv 2 \pi T/a = 3,4,5,10$ of the Unruh length $\beta_U$, Eq.~\eq{eq_T_U}, to the thermal length $\beta_T = 1/T$. The starting point of each cyclic path, $(z,\tau)_i = (z_i,0)$, with $z_i/\beta_U = -1,-1/2, \dots, 1$, is denoted by a hollow circle. The position of the Rindler horizon, collapsed under the Wick rotation to a point~\eq{eq_Rindler_horizon_E}, is denoted by the green star in each plot.
}
\label{fig_acceleration_paths}
\end{figure}

We now move to the Euclidean manifold by performing the Wick rotation to imaginary time, $t \rightarrow \tau = it$. Then, Eq.~\eqref{eq_acc_bc} becomes
\begin{align}
 G_E(\tau_{(1)}, z_{(1)}; x') &= G_E(\tau, z; x'), \nonumber\\
 S_E(\tau_{(1)}, z_{(1)}; x') &= -e^{\alpha S^{0z}} S_E(\tau, z; x'),
 \label{eq_acc_bc_tau}
\end{align}
and Eq.~\eqref{eq_thermal_Feynman} reads, for the case when $\alpha / 2\pi$ is an irrational number,
\begin{subequations}\label{eq_thermal_Feynman_tau}
\begin{align}
 G_E^{(\alpha)}(\tau, z; x') &= \sum_{j = -\infty}^\infty G_E^{\rm vac}(\tau_{(j)}, z_{(j)}; x'), \\
 S_E^{(\alpha)}(\tau, z; x') &= \sum_{j = -\infty}^\infty (-1)^j e^{-j \alpha S^{0z}} S_E^{\rm vac}(\tau_{(j)}, z_{(j)}; x').
\end{align}
\end{subequations}
The case when $\alpha / 2\pi = p/q$ must be treated along the lines summarized in Eqs.~\eqref{eq_thermal_Feynman_rational} (see also discussion in Sec.~\ref{sec:fractal}). In the above, we considered $j \in \Z$ and
\begin{subequations} \label{eq_Euclid_tau_zj_0}
\begin{align}
 \tau_{(j)} &= \tau \cos(j \alpha) - \frac{1}{a} (1 + a z) \sin(j \alpha), \\
 z_{(j)} &= \tau \sin(j \alpha) + \frac{1}{a} (1 + a z) \cos(j \alpha) - \frac{1}{a}.
\end{align}
\end{subequations}
For the fields, the accelerated KMS conditions suggest the identification of the fields at the points:
\begin{subequations}\label{eq_identification}
\begin{align}
	\phi(\tau_{(j)},{\bs x}_\|,z_{(j)}) & = \phi(\tau,{\bs x}_\|,z)\,, \\ 
	\psi(\tau_{(j)},{\bs x}_\|,z_{(j)}) & = (-1)^j e^{j \alpha S^{0z}} \psi(\tau,{\bs x}_\|,z)\,, 
\end{align}	
\end{subequations}
where the identified coordinates $(\tau_{(j)},z_{(j)})$ in the longitudinal plane are given by Eq.~\eq{eq_Euclid_tau_zj_0} and ${\bs x}_\| = (x,y)$ are the transverse coordinates which are unconstrained by acceleration. While the sums of the form~\eq{eq_thermal_Feynman} may formally be divergent, the modified conditions~\eq{eq_Euclid_tau_zj_0} and \eq{eq_identification} give a finite solution to the accelerated KMS relations. The points identified with the accelerated KMS condition~\eq{eq_Euclid_tau_zj_0} are illustrated in Fig.~\ref{fig_acceleration_paths}.

\section{Geometrical meaning of the accelerated KMS relation in imaginary-time formalism} 

It is convenient, for a moment, to define a translationally shifted spatial coordinate, ${\mathsf z} = z + 1/a$, and rewrite Eq.~\eq{eq_Euclid_tau_zj_0} in the very simple and suggestive form:
\begin{align}
 \tau_{(j)} &= \tau \cos(j \alpha) - {\mathsf z} \sin(j \alpha), \nonumber\\
 {\mathsf z}_{(j)} &= \tau \sin(j \alpha) + {\mathsf z} \cos(j \alpha).
 \label{eq_Minkowski_tauzj}
\end{align}
In the shifted coordinates, the condition \eqref{eq_Rindler_horizon_M} for the Rindler horizon becomes $a^2(\mathsf{z}^2 + \tau^2) = 0$, which is solved by
\begin{align}
    \tau = {\mathsf z} = 0 \qquad \Leftrightarrow \qquad \tau = 0, \quad z = - \frac{1}{a}\,.
\label{eq_Rindler_horizon_E}
\end{align}
Thus, we arrive at the following beautiful conclusion: in the Euclidean spacetime of the imaginary-time formalism, the Rindler horizon~\eq{eq_Rindler_horizon_M} shrinks to a single point~\eq{eq_Rindler_horizon_E}. Thus, the accelerated KMS condition corresponds to the identification of all points obtained by the discrete rotation of the space around the Euclidean Rindler horizon point $(\tau, z) = (0, -1/a)$ with the unit rotation angle defined by the reference thermal acceleration $\alpha = a/T$.

\begin{figure}[!ht]
\begin{center}
\includegraphics[width=0.4\textwidth,clip=true]{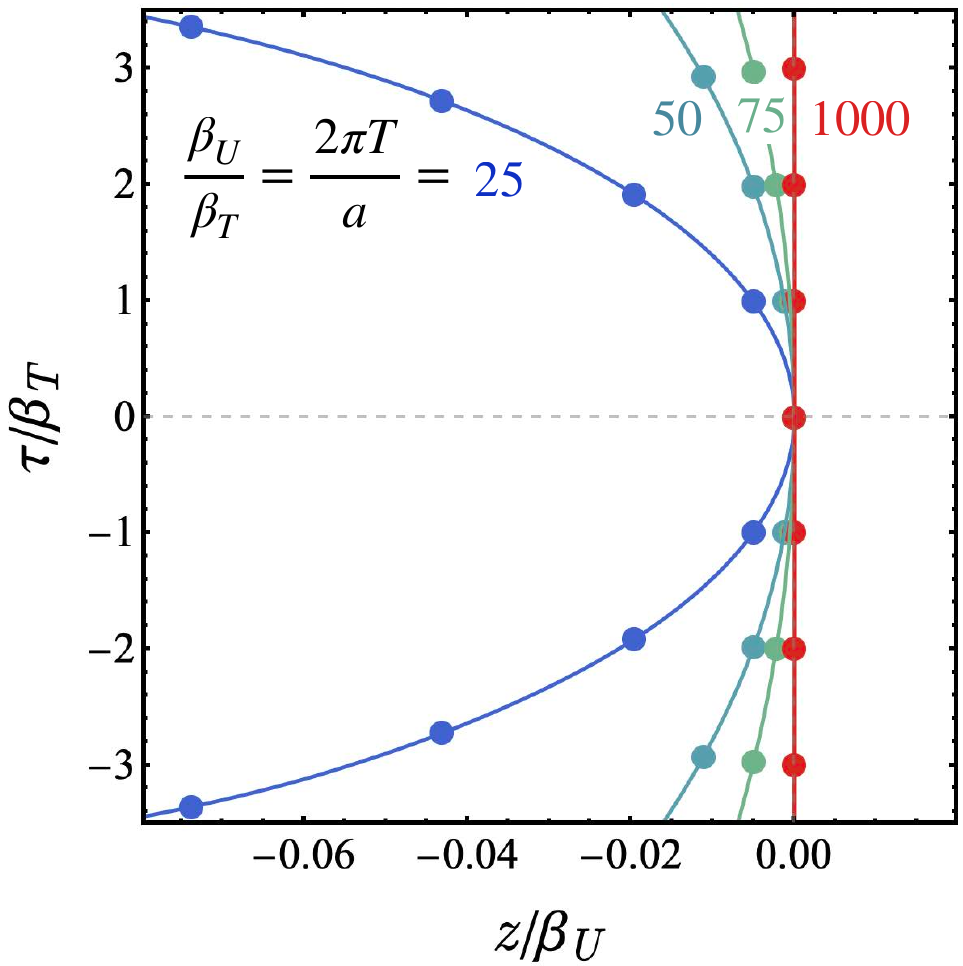}
\end{center}
\caption{
The sets of points in the ($\tau,z$) plane which are identified by our circular KMS condition~\eq{eq_Minkowski_tauzj} with the origin $(\tau, z) = (0,0)$ in a thermally equilibrated system which experiences a uniform acceleration $a$ along the $z$ axis. The color distinguishes different acceleration strength marked by different Unruh lengths $\beta_U = 2 \pi/|a|$. At vanishing acceleration ($\beta_U/\beta_T \to \pm \infty$), condition~\eq{eq_Minkowski_tauzj} reduces to the standard thermodynamic requirement of compactification of imaginary time $\tau$ to a circle with the length $\beta_T = 1/T$, while the Euclidean Rindler horizon moves to (minus) spatial infinity. In the figure, each set of points, corresponding to various ratios $\beta_U/\beta_T$, is connected by a smooth line to guide the eye.}
\label{fig_limit}
\end{figure}

Our accelerated KMS condition, given in Eqs.~\eq{eq_Euclid_tau_zj_0} and \eq{eq_identification}, recovers the usual finite-temperature KMS condition in the limit of vanishing acceleration. Figure~\ref{fig_limit} demonstrates that in this limit,with $\alpha = a/T \to 0$, the proposed KMS-type condition~\eq{eq_Minkowski_tzj} for the acceleration is reduced to the standard finite-temperature KMS-boundary condition~\cite{Kapusta:2006pm} for which imaginary time $\tau$ is compactified to a circle of the length $\beta_T \equiv 1/T$ with the points $(\tau,{\bs x})$ and $(\tau + \beta_T n,{\bs x})$, $n\in \Z$, identified. 

At the critical acceleration $\alpha = 2\pi n$ (with $n \in \Z$), when the background temperature $T$ equals to (an integer multiple of) the Unruh temperature~\eq{eq_T_U}, the accelerated KMS conditions~\eq{eq_Euclid_tau_zj_0} do not constrain the system anymore, $\tau_{(j)} = \tau$ and $z_{(j)} = z$, so that the system becomes equivalent to a zero-temperature system in non-accelerated flat Minkowski spacetime. This property, for $\alpha = 2 \pi$, has been observed in Refs.~\cite{Prokhorov:2019cik,Prokhorov:2019sss,Becattini:2020qol,Palermo:2021hlf}.

In the situation where $2\pi / \alpha = \beta_U / \beta_T = n$ is an integer number, the accelerated state at finite temperature can be implemented in Euclidean space by imposing periodicity with respect to a specific set of points that form a regular polygon with $n$ vertices located on the circle of radius $\tau^2 + z^2$. This is particularly convenient for lattice simulations since the Euclidean action remains the standard one, allowing accelerated systems to be modeled in the imaginary-time path integral formalism without encountering the infamous sign problem.

\section{KMS relations in Rindler coordinates}

In the Minkowski Lorentz frame that we considered so far, the accelerating KMS conditions~\eq{eq_Euclid_tau_zj_0} and \eq{eq_identification} do not correspond to a boundary condition (as one would naively expect from the KMS condition in thermal field theory) but rather to a bulk condition: instead of relating the points at the boundary of the imaginary-time Euclidean system, the accelerated KMS relations give us the identification of the spacetime points in its interior. 

While seemingly non-trivial in the form written in Eq.~\eqref{eq_Minkowski_tzj}, the displacements implied by the KMS relation correspond to the usual translation of the proper time (rapidity) coordinate $\eta$ when employing the Rindler coordinates,
\begin{equation}
 at = e^{\zeta} \sinh(a \eta), \quad 
 1 + az = e^{\zeta} \cosh(a \eta).
 \label{eq_Rindler_coordinates}
\end{equation}
It is easy to see that 
\begin{subequations}
\begin{align}
 at_{(j)} & = e^\zeta \sinh(a \eta + i j \alpha), \\ 
 1 + a z_{(j)} & = e^\zeta \cosh(a \eta + i j \alpha),
\end{align}
\end{subequations}
\!\!which implies that 
\begin{equation}
 \eta_{(j)} = \eta + i j \beta_T, \qquad 
 \zeta_{(j)} = \zeta,
 \label{eq_Rindler_KMS}
\end{equation}
in a seemingly perfect agreement with the usual KMS relation \eqref{eq_standard_KMS} for static systems in Minkowski. However, there is also an unusual particularity of the KMS conditions~\eq{eq_Rindler_KMS} in the Rindler coordinates~\eq{eq_Rindler_coordinates}.

The first relation in Eq.~\eq{eq_Rindler_KMS} suggests that the Wick rotation of the Minkowski time $t = - i \tau$ should be supplemented with the Wick rotation of the proper time in the accelerated frame $\eta = - i \theta/a$, where $\theta$ is the imaginary rapidity.\footnote{Named in analogy with the rapidity coordinate $\psi \equiv a \eta$.} Then, the relation~\eq{eq_Rindler_coordinates} in the imaginary (both Minkowski and Rindler) time becomes as follows:
\begin{equation}
 a \tau = e^{\zeta} \sin\theta, \quad 
 1 + az = e^{\zeta} \cos\theta,
 \label{eq_Rindler_coordinates_Im} 
\end{equation}
which shows that the imaginary rapidity becomes an imaginary coordinate with the Euclidean Rindler KMS condition~\eq{eq_Rindler_KMS}:
\begin{equation}
 \theta_{(j)} = \theta - j \alpha, \qquad \zeta_{(j)} = \zeta, \qquad j \in \Z\,.
 \label{eq_Rindler_KMS_Im}
\end{equation}
Curiously, under the Wick transform, the rapidity becomes a cyclic compact variable, $0 \leqslant \theta < 2\pi$, on which the imaginary-time condition~\eq{eq_Rindler_KMS_Im} imposes the additional periodicity with the period equal to the thermal acceleration~$\alpha$. Expectedly, at $\alpha = 2 \pi$ (or, equivalently, at $T = T_U$), the boundary condition~\eq{eq_Rindler_KMS_Im} becomes trivial.

The boundary conditions~\eq{eq_Rindler_KMS_Im}, characterized by the doubly-periodic imaginary rapidity coordinate $\theta$, with periodicities $\theta \to \theta + 2 \pi$ and $\theta \to \theta + \alpha$ (for $0 \leqslant \alpha < 2\pi$), can be easily implemented in lattice simulations. Notice that this double periodicity has a strong resemblance to the observation of Refs.~\cite{Prokhorov:2019hif,Prokhorov:2019yft,Zakharov:2020ked} that the Euclidean Rindler space can be identified with the space of the cosmic string which possesses a conical singularity with the angular deficit $\Delta \varphi = 2\pi - \alpha$ \cite{Dowker:1987pk,Linet:1994tz}.

The KMS periodicity~\eq{eq_Rindler_KMS_Im} of the compact imaginary rapidity $\theta$ is formally sensitive to the rationality of the normalized thermal acceleration $\alpha/(2 \pi)$. Obviously, for $\alpha = 2\pi p/q$, where $p<q$ are nonvanishing irreducible integer numbers, the interplay of the two periodicities will correspond to the single period $\theta \to \theta + 2\pi/q$.

Interestingly, the sensitivity of an effect to the denominator $q$ (and not to the numerator $p$) of a relevant parameter is a signature of the fractal nature of the effect. Such fractality is noted, for example, in particle systems subjected to imaginary rotation implemented via rotwisted boundary conditions~\cite{Chernodub:2020qah,Chen:2022smf,Ambrus:2023bid}, which leads, in turn, to the appearance of ``ninionic'' deformation of particle statistics~\cite{Chernodub:2022qlz}. The suggested fractality of acceleration in imaginary formalism is not surprising given the conceptual similarity of acceleration and rotation with imaginary angular frequency~\cite{Becattini:2020qol,Palermo:2021hlf}. Below, we will show that, despite the fractal property of the system, the KMS boundary condition~\eq{eq_Rindler_KMS_Im} in Euclidean Rindler space correctly reproduces results for accelerated particle systems.

\section{Energy-momentum tensor with the accelerated KMS conditions}

Now let us come back to the Wick-rotated Minkowski spacetime and verify how the modified KMS conditions for the fields, Eqs.~\eq{eq_Euclid_tau_zj_0} and \eq{eq_identification}, and related solutions for their two-point functions~\eq{eq_thermal_Feynman_tau}, can recover the known results in field theories under acceleration. To this end, we start from a non-minimally coupled scalar field theory with the Lagrangian~\cite{Callan1970,Frolov1987,Becattini:2015nva}
\begin{align}
	{\mathcal L}_\xi = \frac{1}{2} \partial_\mu \phi \partial^\mu \phi 
     - 2 \xi \partial_\mu \left( \phi \partial^\mu \phi \right),
	\label{eq_L_xi}
\end{align}
possessing the following energy-mo\-men\-tum tensor:
\begin{align}
 \Theta^\xi_{\mu\nu} = (1 - 2\xi) \partial_\mu \phi \partial_\nu \phi & - 2\xi \phi \partial_\mu \partial_\nu \phi \nonumber\\
 & - \frac{1}{2} (1- 4\xi) \delta_{\mu\nu} \partial_\lambda \phi \partial_\lambda \phi, \label{eq_Theta_xi} 
\end{align}
where the values $\xi = 0$ and $\xi = 1/6$ of the coupling parameter correspond to the canonical 
and conformal energy-momentum tensors, respectively. In terms of the Euclidean Green's function, $\Theta^\xi_{\mu\nu}$ can be written as 
\begin{multline}
 \Theta^\xi_{\mu\nu} = \lim_{x'\rightarrow x} \left[(1 - 2\xi) \partial_{(\mu} \partial_{\nu')}  - \tfrac{1}{2} (1 - 4\xi) \delta_{\mu\nu} \partial_\lambda \partial_{\lambda'}\right.\\
 \left. - \xi (\partial_\mu \partial_\nu + \partial_{\mu'} \partial_{\nu'})\right] \Delta G^{(\alpha)}_E(x,x'),
\end{multline}
where $\Delta G^{(\alpha)}_E(x,x') = G^{(\alpha)}_E(x,x') - G_E^{\rm vac}(x,x')$ represents the thermal part of the Green's function. For the Dirac field, $\Theta_{\mu\nu} = \frac{1}{2} \bar{\psi} \gamma^E_\mu \overleftrightarrow{\partial_\nu} \psi$ can be computed from the Euclidean two-point function $S^{(\alpha)}_E(x,x')$ via
\begin{equation}
 \Theta_{\mu\nu} = \frac{1}{2} \lim_{x' \rightarrow x} {\rm tr} [\gamma^E_\mu (\partial_\nu - \partial_{\nu'}) \Delta S^{(\alpha)}_E].
\end{equation}

The vacuum propagators satisfying $\Box G^{\rm vac}_E(x,x') = \gamma^E_\mu \partial_\mu S^{\rm vac}_E(x,x')  = \delta^4(x-x')$ are given by
\begin{align}
 G^{\rm vac}_E(\Delta x) &= \frac{1}{4\pi^2 \Delta X^2}, \label{eq_G_E}\\
 S_E^{\rm vac}(\Delta x) &= \gamma^E_\mu \partial_\mu G^{\rm vac}_E(\Delta x) = -\frac{\gamma^E_\mu \partial_\mu}{2\pi^2 \Delta X^4},
\end{align}
with $\Delta X^2 = (\Delta \tau)^2 + (\Delta {\bs x})^2$. Using Eq.~\eqref{eq_thermal_Feynman_tau}, the thermal expectation values of the normal-ordered energy-momentum operator can be obtained in the case of the Klein-Gordon field as:
\begin{multline}
 \Theta^{\mu\nu}_\xi(x) = \sum_{j \neq 0}^\infty \frac{1}{4\pi^2 \Delta X_{(j)}^4} \left[(1 - 2\xi) (R^{(j)}_{\mu\nu} + R^{(j)}_{\nu\mu}) \right. \\
 \left. - \delta_{\mu\nu}(1 - 4\xi) R^{(j)}_{\lambda\lambda} + 2\xi(R^{(j)}_{\nu\lambda} R^{(j)}_{\mu\lambda} + \delta_{\mu\nu}) \right] \\
 - \sum_{j \neq 0}  \frac{\Delta x^{(j)}_\lambda \Delta x^{(j)}_\kappa}{\pi^2 \Delta X^6_{(j)}} \left[(1 - 2\xi) (\delta_{\mu\lambda} R^{(j)}_{\nu\kappa} + \delta_{\nu\lambda} R^{(j)}_{\mu\kappa}) \right. \\
 \left. - \delta_{\mu\nu}(1 - 4\xi) R^{(j)}_{\lambda\kappa} + 2\xi (R^{(j)}_{\mu\lambda} R^{(j)}_{\nu\kappa} + \delta_{\mu\lambda} \delta_{\nu\kappa})\right],
\end{multline}
where $\Delta X^2_{(j)} = \frac{4}{a^2} \sin^2 \frac{j  \alpha}{2} [(a\tau)^2 + (1 + az)^2]$ and $R^{(j)}_{\mu\nu} \equiv \partial_\mu \Delta x^{(j)}_\nu$ is given by 
\begin{equation}
 R^{(j)}_{\mu\nu} = \begin{pmatrix}
     \cos (j \alpha) & 0 & 0 & \sin (j \alpha) \\
     0 & 1 & 0 & 0\\
     0 & 0 & 1 & 0 \\
     -\sin (j \alpha) & 0 & 0 & \cos (j \alpha)
 \end{pmatrix},
\end{equation}
such that $R^{(j)}_{\mu\lambda} R^{(j)}_{\nu\lambda} = \delta_{\mu\nu}$.
For the Dirac field, we find
\begin{multline}
 \Theta_{\mu\nu} = -\sum_{j \neq 0} \frac{(-1)^j}{\pi^2 } \left[\delta_{\mu\lambda} \cos \tfrac{j \alpha}{2} + \left(\delta_{\mu 0} \delta_{\lambda 3} - \delta_{\mu 3} \delta_{\lambda 0} \right) \sin \tfrac{j \alpha}{2} \right] \\
 \times \left[\frac{R_{\nu \lambda}^{(j)} + \delta_{\nu \lambda}}{\Delta X^4_{(j)}} - 
 \frac{4 \Delta X^{(j)}_{\lambda}}{\Delta X_{(j)}^6} (R^{(j)}_{\nu\kappa} + \delta_{\nu\kappa})\Delta X^{(j)}_\kappa\right]. 
\end{multline}

Taking advantage of the relation $(R^{(j)}_{\nu\kappa} + \delta_{\nu\kappa}) \Delta X^{(j)}_\kappa = -\frac{2}{a} \sin(j \alpha) [(1 + a z) \delta_{\nu 0} - a\tau \delta_{\nu 3}]$ and after switching back to the real time $t$, we find
\begin{equation}
 \Theta^{\mu\nu} = \mathcal{E} u^\mu u^\nu - \mathcal{P} \Delta^{\mu\nu} + \pi^{\mu\nu},
 \label{eq_Theta_munu}
\end{equation}
with $\mathcal{E}$, $\mathcal{P}$, and $u^\mu$ being the energy density, isotropic pressure, and the fluid four-velocity~\eq{eq_u}, respectively. The shear-stress tensor $\pi^{\mu\nu}$ is by construction traceless, symmetric and orthogonal to $u^\mu$, discriminating between the energy-momentum tensors in {\it classical}~\eq{eq_Tmunu} and {\it quantum}~\eq{eq_Theta_munu} fluids. Due to the symmetries of the problem, its tensor structure is fixed as 
\begin{align}
\pi^{\mu\nu} = \frac{\pi_s}{2} \left(\Delta^{\mu\nu} - \frac{3 \alpha^\mu \alpha^\nu}{\alpha^\lambda \alpha_\lambda}\right)\,,	
	\label{eq_pi_munu}
\end{align}
with $\alpha^\mu(x)$ being the local thermal acceleration~\eq{eq_a}, such that the shear coefficient $\pi_s$ is the only degree of freedom of $\pi^{\mu\nu}$ in Eq.~\eq{eq_pi_munu}.
In the scalar case, we find for the components of \eq{eq_Theta_munu}:
\begin{align}
 \mathcal{E}_\xi &= \frac{3 [\alpha T(x)]^4}{16 \pi^2} \left[G_4(\alpha) - 4\xi G_2(\alpha)\right], \nonumber\\
 \mathcal{P}_\xi &= \frac{[\alpha T(x)^4]}{16 \pi^2} \left[G_4(\alpha) -\frac{4}{3}\left(1 - 3\xi\right) G_2(\alpha)\right], \nonumber\\
 \pi_s^\xi &= \frac{[\alpha T(x)]^4}{12 \pi^2} (1 - 6\xi) G_2(\alpha),
 \label{eq_KG_res}
\end{align}
with $G_n(\alpha) = \sum_{j = 1}^\infty [\sin(j\alpha / 2)]^{-n}$, in complete agreement with the results in Ref.~\cite{Becattini:2020qol}. Formally, $G_n$ diverges, however its value can be obtained from its analytical continuation to imaginary acceleration $a = i \phi$, $\widetilde{G}_n(\beta_T \phi) = i^n G_n(i \beta_T \phi)$. The sum can be evaluated, in a certain domain around $\beta_T \phi > 0$~\cite{Becattini:2020qol}, to:
\begin{align}
 \widetilde{G}_2(\beta_T \phi) & = \frac{2\pi^2}{3 \beta_T^2 \phi^2} - \frac{2}{\beta_T \phi}
 + \frac{1}{6}, \nonumber\\
 \widetilde{G}_4(\beta_T\phi) &= 
 \frac{8\pi^4}{45 \beta_T^4 \phi^4} - \frac{4\pi^2}{9\beta_T^2\phi^2} + \frac{4}{3 \beta_T \phi} - \frac{11}{90}.
\end{align}
Substituting now $G_n(\alpha) = {\rm Re} [i^{-n} \widetilde{G}_n(\beta_T \phi) \rfloor_{\phi \rightarrow -i a}]$ into Eq.~\eqref{eq_KG_res} gives Eq.~\eqref{eq_E} for the conformal coupling $\xi = 1/6$. For minimal coupling $\xi = 0$ or a generic non-conformal coupling $\xi \neq 1/6$, we recover the results of Refs.~\cite{Becattini:2020qol,Zakharov2020}.

In the case of the Dirac field, one can easily check that $\mathcal{E}_D = 3\mathcal{P}_D$ and $\pi_D^s = 0$, while
\begin{equation}
 \mathcal{P}_D = \frac{[\alpha T(x)]^4}{4\pi^2} S_4( \alpha),
\end{equation}
with $S_n(\alpha) = -\sum_{j = 1}^\infty (-1)^j \cos(j\alpha/2) / [\sin(j\alpha / 2)]^n \rightarrow \widetilde{S}_n(\beta_T\phi)$ $\equiv i^n S_n(i \beta_T \phi) = - \sum_{j=1}^\infty (-1)^j \cosh(j\beta_T \phi/2) / [\sinh(j\beta_T\phi/2)]^n$, which agrees with the results obtained in Ref.~\cite{Palermo:2021hlf}.

Finally, let us also illustrate the practical functionality of the accelerating KMS boundary conditions~\eq{eq_Rindler_KMS_Im} formulated in the imaginary-rapidity Rindler space~\eq{eq_Rindler_coordinates_Im}. For simplicity, we calculate the fluctuations of the scalar field $\langle \phi^2 \rangle$ using point-splitting and noticing that the same method can be used to calculate also other quantities.

When expressed with respect to Rindler coordinates $X = (\theta/a, \mathbf{x}_\perp, \zeta)$,
the Euclidean vacuum two-point function $G_{E,R}^{\rm vac}(X,X')$ given in Eq.~\eqref{eq_G_E} reads as follows:
\begin{equation}
 G_{\rm E,R}^{\rm vac} = \frac{1}{4\pi^2} \left[ 
 \frac{2}{a^2} e^{\zeta + \zeta'} (\cosh \Delta \zeta - \cos \Delta \theta) + \Delta {\bs x}_\perp^2\right]^{-1}.
\end{equation}
The KMS condition~\eq{eq_Rindler_KMS_Im} implies that the Euclidean two-point function under acceleration satisfies $G^{(\alpha)}_{\rm E,R} = \sum_{j \in \Z} G^{\rm vac}_{\rm E,R} (\Delta \theta + j \alpha)$, where we consider vanishing spatial distance between the points: $\zeta' \to \zeta$ and $\mathbf{x}_\perp' \to \mathbf{x}_\perp$. Subtracting the vacuum ($j=0$) term that diverges in the $\Delta X \to 0$ limit, we get for the scalar fluctuations:
\begin{align}
    \langle \phi^2 \rangle & = \lim_{\Delta \theta \to 0} \bigl[G^{(\alpha)}_{\rm E,R} (\Delta \theta) - G^{\rm vac}_{\rm E,R} (\Delta \theta)\bigr]  
    \label{eq_phi2}\\
    & 
    = \frac{a^2 e^{-2\zeta}}{8\pi^2} G_2(\alpha) = \frac{T^2(x)}{12} - \frac{a^2(x)}{48\pi^2}\,, 
    \quad 
    0 \leqslant a \leqslant 2 \pi T\,,\nonumber
\end{align}
which agrees with the known result~\cite{Becattini:2020qol,Diakonov:2023hzg}. 

\section{Fractalization of thermodynamics}\label{sec:fractal}

Let us consider the case when $\alpha / 2\pi$ is a rational number, represented as the irreducible fraction $p / q$. Then, the functions $G_n(\alpha) \rightarrow G_n^{(p,q)}(\alpha) = \frac{1}{2} \sum_{j = 1}^{q-1} [\sin(\pi j p / q)]^{-n}$ are regular and evaluate in the relevant $n = 2$ and $n = 4$ cases to
\begin{equation}
 G_2^{(p,q)} = \frac{q^2 - 1}{6}, \quad 
 G_4^{(p,q)} = \frac{q^4 + 10q^2 -11}{90}.
\end{equation}
The above results are independent of the numerator $p$ of the irreducible fraction. The quadratic field fluctuations, shear stress coefficient $\pi_s$, energy density, and pressure reduce to
\begin{subequations}
\begin{align}
 \langle\phi^2\rangle^{(p,q)} &= \frac{[\alpha T(x)]^2}{48\pi^2} (q^2 - 1), \\
 \mathcal{E}_\xi^{(p,q)} &= \frac{[\alpha T(x)]^4}{480\pi^2} (q^2 - 1)(q^2 + 11 - 60\xi), \\
 \mathcal{P}_\xi^{(p,q)} &= \frac{[\alpha T(x)]^4}{1440\pi^2} (q^2 - 1)(q^2 - 9 + 60\xi), \\
 \pi_{s;\xi}^{(p,q)} &= \frac{[\alpha T(x)]^4}{72\pi^2} (1 - 6\xi) (q^2 - 1),
\end{align}
\end{subequations}
manifestly vanishing when $q^2 = 1$, i.e. for $\alpha = 2\pi$.

In the case of the Dirac field, we have $S_n(\alpha) \rightarrow S_n^{(p,q)} = -\frac{1}{2} \sum_{j = 1}^{q-1} (-1)^j \cos(\pi j p / q) / [\sin(\pi j p / q)]^n$. For the case $n = 4$, the relation $(-1)^{q-j} \cos[\pi(q-j)p / q] = (-1)^{j + p + q} \cos(\pi j p / q)$ implies that $S_4^{(p,q)}$ vanishes when $p + q$ is an odd number. This happens whenever $q$ is an even number in order to maintain the fraction $p / q$ irreducible. When $q$ is odd, $S_4^{(p,q)}$ vanishes for all even values of $p$. When both $p$ and $q$ are odd, $S_4^{(p,q)}$ can be computed analytically and the final result can be summarized as
\begin{equation}
 S_4^{(p,q)} = 
     \frac{7 q^2 + 17}{720} (q^2 - 1) \times 
     \frac{1 + (-1)^{p+q}}{2}.
\end{equation}
The fermion pressure becomes 
\begin{equation}
 \mathcal{P}^{(p,q)}_D = \frac{[\alpha T(x)]^4}{2880\pi^2} (q^2 -1)(7q^2 + 17) \frac{1 + (-1)^{p+q}}{2}.
\end{equation}

\section{Conclusions} 

In this paper, we derived the KMS relation for bosonic and fermionic quantum systems at finite temperature under uniform acceleration. In Wick-rotated Minkowski spacetime, the uniform acceleration requires the identification~\eq{eq_Euclid_tau_zj_0} of the points in the {\it bulk} of the system along the discrete points lying on circular orbits~\eq{eq_identification} about the Rindler horizon, which shrinks to a point~\eq{eq_Rindler_horizon_E} under the Wick rotation. In the Wick-rotated Rindler coordinates, the KMS relations reduce to standard (anti-)periodic boundary conditions in terms of the imaginary rapidity coordinates. To illustrate the effectiveness of the method, we considered the quantum thermal distributions of massless scalar and Dirac particles under acceleration and found perfect agreement with results previously derived in the literature. 

Our work paves the way to systematic explorations of the influence of the kinematic state of a system on its global equilibrium thermodynamic properties. Our paper equips us with a rigorously formulated method in imaginary-time formalism which allows us to construct the ground state of a field theory in thermal equilibrium in a uniformly accelerating frame, opening, in particular, a way for first-principle lattice simulations of accelerated systems. 

\section*{Acknowledgements} The authors are indebted to our colleagues: Dr. Andrea Palermo, Dr. Matteo Buzzegoli and Prof. Francesco Becattini for carefully reading our work and for drawing our attention to (some of) the errors in previous versions of this paper. This work is supported by the European Union - NextGenerationEU through the grant No. 760079/23.05.2023, funded by the Romanian ministry of research, innovation and digitalization through Romania’s National Recovery and Resilience Plan, call no. PNRR-III-C9-2022-I8.

\bibliographystyle{elsarticle-num}
\bibliography{plasma}

\end{document}